\documentclass{aa}
\usepackage{txfonts}
\usepackage{natbib}
\usepackage{graphicx}

\bibpunct[, ]{(}{)}{;}{a}{}{,}

\newcommand{\fifps}[2]{\centering\resizebox{#1}{!}{\includegraphics{#2}}}
\newcommand{\figps}[1]{\resizebox{\hsize}{!}{\rotatebox{0}{\includegraphics{#1}}}}

\newcommand{\filps}[1]{\resizebox{\hsize}{!}{\rotatebox{-90}{\includegraphics{#1}}}}

\def\llm{{\sc LLModels}}
\def\atlas{{\sc atlas9}}
\def\tef{T_{\rm eff}}
\def\tauros{\tau_{\rm ross}}
\def\kms{km\,s$^{-1}$}

\begin{document}

\title{Stellar model atmospheres with magnetic line blanketing}

\author{O. Kochukhov\inst{1}\thanks{\emph{Present address:} 
NORDITA, Blegdamsvej 17, 2100 Copenhagen \O, Denmark} 
\and S. Khan\inst{1,2} \and D. Shulyak\inst{1,2}}

\offprints{O. Kochukhov, \\ \email{kochukhov@astro.univie.ac.at}}

\institute{ Institut f\"ur Astronomie, Universit\"at Wien, T\"urkenschanzstra{\ss}e 17, 1180 Wien, Austria 
\and Tavrian National University, Yaltinskaya 4, 95007 Simferopol, Crimea, Ukraine }

\date{Received / Accepted }

\abstract{
Model atmospheres of A and B stars are computed taking into account magnetic
line blanketing. These calculations are based on the new stellar model
atmosphere code \llm\ which implements direct treatment of the opacities due to
the bound-bound transitions and ensures an accurate and detailed description of
the line absorption. The anomalous Zeeman effect was calculated for the field
strengths between 1 and 40\,kG and a field vector perpendicular to the line of
sight. The model structure, high-resolution energy distribution, photometric
colors, metallic line spectra and the hydrogen Balmer line profiles are computed for magnetic stars
with different metallicities and are discussed with respect to those of
non-magnetic reference models. The magnetically enhanced line blanketing
changes the atmospheric structure and leads to a redistribution of energy in the
stellar spectrum. The most noticeable feature in the optical region is the
appearance of the 5200\,\AA\ depression. However, this effect is
prominent only in cool A stars and disappears for higher effective
temperatures. The presence of a magnetic field produces opposite variation of
the flux distribution in the optical and UV region. A deficiency of the UV flux
is found for the whole range of considered effective temperatures, whereas the
``null wavelength'' where flux remains unchanged shifts towards the
shorter wavelengths for higher temperatures.  
}

\maketitle

\section{Introduction}

Magnetic chemically peculiar (CP) stars are the upper and middle main sequence
stars characterized by anomalous chemical abundances and an unusual distribution
of energy in their spectra. $\tef$ of magnetic CP stars ranges from 6500\,K to
about 25\,000\,K, whereas the typical magnetic field strengths observed in
their atmospheres span from a few hundred gauss to \mbox{$\sim30$\,kG}. A strong
magnetic field is expected to have an important effect on the atmospheres and
spectra of CP stars. Whereas several detailed line profile studies
\citep[e.g.][]{kochukhov} have included the full treatment of the Zeeman effect in
calculation of the Stokes parameters of individual spectral lines and short
wavelength regions, an overall influence of the magnetic absorption on the
structure of the model atmospheres and flux distributions of CP stars remained
relatively unexplored. 

Several characteristic features in the energy distribution of magnetic CP stars
are suspected to be a result of the enhanced line blanketing due to the magnetic
intensification of spectral lines. For instance, \citet{kodaira} found several
flux depressions in the visual spectrum of CP stars, and \citet{leckrone1}
showed that the UV flux of CP stars is depressed compared to that of normal
stars with a similar flux distribution in the optical region. These possible
manifestations of the magnetic line blanketing emphasize a necessity to
consider a magnetic field in the model atmosphere calculation of CP stars. 

In general, a magnetic field influences the energy transport, hydrostatic
equilibrium, diffusion processes and the formation of spectral lines. Previous
studies \citep{stepien,muthsam,carpenter,leblanc,valyavin} have made attempts
to model some of these factors. However, due to a limitation of computer
resources, it was impossible to fully account for the Zeeman effect on the line
absorption, except in the modelling of the pure hydrogen atmospheres of magnetic 
white dwarfs \citep{wikra}. In the early model atmosphere calculations 
for non-degenerate stars magnetic splitting was
treated very approximately by introducing a pseudo-microturbulent velocity
which affects all spectral lines in the same way \citep{muthsam} or by adopting
a normal Zeeman triplet pattern with the same effective Land\'{e} factors
\citep{carpenter} for all lines. Recent investigation by \citet{stift} demonstrated that
the magnetic intensification of spectral lines depends primarily on the parameters
of the anomalous Zeeman splitting pattern, in particular on the number of Zeeman
components. In the light of these results it becomes clear that previous
attempts to simulate magnetic line blanketing by an enhanced microturbulence or
by using a simple triplet pattern are insufficient. 

The primary aim of our paper is to introduce a realistic calculation of the
anomalous Zeeman effect in the classical 1-D models of stellar atmospheres and
to investigate the resulting effects on the model structure, energy distribution
and other common observables. We calculate a grid of model atmospheres of A and
B stars assuming a horizontal magnetic field and exploring a $\tef$ range relevant
for CP stars. The paper is organized as follows. In Sect.\,\ref{techniques} we
describe our model atmosphere code and numerical implementation of the Zeeman
effect in the line opacity calculation. Sect.\,\ref{results} presents numerical
results and Sect.\,\ref{discussion} summarizes our work and compares the
observed and predicted anomalies in the shape and variability of the flux
distribution of magnetic CP stars.

\section{Calculation of magnetic model atmospheres}
\label{techniques}
\subsection{The stellar model atmosphere code LLModels}

\begin{table*}
\centering
\caption{Description of the spectral line lists used in the line blanketing
calculations for different effective temperature $\tef$ and metallicity
$[M/H]\equiv\lg(N_{\rm metals}/N_{\rm H}) -\lg(N_{\rm metals}/N_{\rm H})_\odot$. 
$N_{\rm pres}$ is the number of spectral lines retained by the 
preselection procedure, $N_{\rm Land\acute{e}\,\,pres}$ -- the number of spectral lines with known
Land\'e factors, $N_{\rm Land\acute{e}\,LS}$ -- the number of spectral lines of
light elements for which Land\'e factors were calculated
assuming LS coupling, $N_{\rm Land\acute{e}\,triplet}$ -- the number of spectral
lines of light and heavy elements without a proper term designation in VALD which were
splitted assuming a classical Zeeman triplet pattern with the effective Land\'e factor
1.2, $N_{\rm total}$ -- the total number of the Zeeman components included in the
line blanketing calculation.}
\begin{tabular}{lllllll}
\hline\hline
$\tef$ & $[M/H]$ & $N_{\rm pres}$ & $N_{\rm Land\acute{e}\,pres}$ 
& $N_{\rm Land\acute{e}\,LS}$ & $N_{\rm Land\acute{e}\,triplet}$ & $N_{\rm total}$ \\
\hline
  8000    & 0.0 & 333\,940 & 305\,109 & 16\,968 & 11\,863 &  5\,910\,394 \\
          & 0.5 & 491\,340 & 455\,591 & 20\,371 & 15\,378 &  8\,774\,132 \\
          & 1.0 & 719\,988 & 676\,889 & 23\,695 & 19\,404 & 12\,983\,819 \\
\hline
11\,000   & 0.0 & 284\,361 & 263\,739 & 14\,462 &  6\,160 &  5\,187\,013 \\
          & 0.5 & 418\,093 & 390\,641 & 18\,973 &  8\,479 &  7\,707\,542 \\
          & 1.0 & 624\,273 & 588\,005 & 24\,285 & 11\,983 & 11\,625\,934 \\
\hline
15\,000   & 0.0 & 346\,912 & 326\,327 & 15\,428 &  5\,157 &  6\,438\,224 \\
          & 0.5 & 519\,902 & 493\,198 & 19\,114 &  7\,590 &  9\,731\,721 \\
          & 1.0 & 785\,203 & 751\,050 & 23\,542 & 10\,611 & 14\,827\,741 \\
\hline
\end{tabular}
\label{number_lines}
\end{table*}

In calculations described in the present paper we employed the stellar model
atmosphere code \llm\ developed by \citet{llmodels2}. This code uses a direct
method, the so-called line-by-line or LL technique, for the line opacity
calculation. Such an approach allowed us to account for the anomalous Zeeman
splitting of spectral lines in the line blanketing calculation and, hence, to
achieve a qualitatively new level of accuracy in modelling the influence of
magnetic field on the stellar atmospheric structure.

\llm\ is the stellar model atmosphere code which aims at modelling the early
and intermediate type stars taking into account their individual chemical
composition and an inhomogeneous vertical distribution of elemental abundances.
The code is based on the modified \atlas\ subroutines \citep{kurucz13} and on
the spectrum synthesis code described by \citet{tsymbal}. The \llm\ code is
written in Fortran\,90 and uses the following general approximations:
\begin{itemize}
\item the plane-parallel geometry is assumed;
\item the Local Thermodynamic Equilibrium (LTE) is used to calculate the atomic
level populations for all chemical species;
\item the stellar atmosphere is assumed to be in a steady state;
\item the radiative equilibrium condition is fulfilled.
\end{itemize}
The main goal of the LL modelling technique is to avoid a simplified or
statistical description of the bound-bound opacity in stellar atmospheres. The
temperature distribution for a given model strongly depends upon the accuracy
of the flux integration over the whole spectral range where the star radiates
significantly. Consequently, to ensure an accurate flux integration, the number
of frequency points must be sufficient to provide a realistic description of
individual spectral lines. In the past limited computer resources and imperfect
numerical techniques were required to solve this problem. The best known
approximations are the Opacity Distribution Function method
\citep[ODF,][]{kurucz1979} and the Opacity Sampling technique
\citep[OS,][]{gustafsson}.

The key idea of the ODF method is to replace the line opacity at a given
wavelength interval $\Delta\lambda$ by a smooth distribution function
$f(\Delta\lambda)$. The ODF tables are calculated for a set of $T-P$ pairs,
chemical compositions (solar or scaled solar) and microturbulent velocities. A
sparse wavelength grid is sufficient for the flux integration and, thus, the method
achieves a substantial saving of computing time. On the other hand, the OS
method relies on statistically random distribution of wavelength points over the
whole spectral range. A sufficient accuracy is considered to be obtained when
the flux integral becomes independent from the choice of the wavelength grid
within a given error.

Due to their statistical nature, both the ODF and the OS methods suffer from
serious shortcomings. For example, in the case of the ODF technique it is
necessary to recalculate the ODF tables each time when new stellar abundances are
used or an input line list is modified due to the magnetic effects. Obviously, this
method is not efficient in the context of the present study because of the
necessity to compute the ODF tables for a large number of magnetic field strengths
and chemical compositions. On the other hand, the accuracy of the OS method
directly depends on the number of frequency points and their distribution over
the considered spectral range. This requires a careful calibration of the
method each time an important input parameter, such as the magnetic field
strength, is modified.

With the LL method we do not use any precalculated opacity tables and, hence, do
not require an excessive hard disk storage space. Furthermore,
no approximations about the line opacity coefficient are being made during the
model calculation. By using a large number of wavelength points ($\approx$
300\,000 -- 500\,000) we provide a detailed description of the line absorption
for all atmospheric depths and achieve a
higher dynamical range in opacity. This allows to reach a required accuracy of the
resulting model atmosphere structure, especially in the upper atmospheric layers.
Due to the efficient numerical algorithms
implemented in \llm, the code is able to compute model atmospheres
with modern PCs in a reasonable amount of time \citep{llmodels2}. 

In all calculations presented here we employed two criteria to control the
convergence of models: the constancy of the total flux {\em and} conservation
of the radiative equilibrium. Both criteria are checked after each iteration
and for each atmospheric depth. If the condition
\begin{equation}
(H_{\rm rad}+H_{\rm conv})-\sigma \tef^4 = 0
\end{equation}
and the total energy balance
\begin{equation}
\int(\alpha_\nu + \ell_\nu) J_\nu d\nu - \int(\alpha_\nu+\ell_\nu) S_\nu d\nu = 0
\end{equation}
are satisfied in all layers with errors of less than 1\%, we assume that a
model has converged. Here the $H_{\rm rad}$ and $H_{\rm conv}$ represent the
radiative and convective fluxes, $\tef$ is the effective temperature,
$\alpha_\nu$ and $\ell_\nu$ are the continuum and line absorption coefficients at
the frequency $\nu$, whereas the $J_\nu$ and $S_\nu$ are the mean intensity and
source functions respectively. We note that the first criterion is not very
sensitive to the temperature variations in the optically thin layers. Is is the
radiative equilibrium condition which becomes more relevant in this case. If any
of these criteria cannot be satisfied, either the initial model parameters were
inappropriate or the model does not converge due to fundamental limitations of
the adopted physical description of the stellar atmospheric processes.

The \llm\ code uses either atomic line lists compiled by \citet{kurucz13} or the 
Vienna Atomic Line Database \citep[VALD,][]{vald}
line list which is converted to a special binary format. In model
atmosphere calculations there is no need to use all available spectral line data
and include very weak lines. Hence, the \llm\ code relies on a preselection
procedure to choose spectral lines contributing significantly to the total
absorption coefficient. The code selects spectral lines for which
${\ell_\nu}/{\alpha_\nu}\geq \varepsilon$, where $\varepsilon$ is the adopted
selection threshold and $\ell_\nu$ and $\alpha_\nu$ are defined above.

\subsection{Line lists}

The magnetic line blanketing was taken into account for all spectral lines except
the hydrogen lines according to the individual anomalous Zeeman splitting pattern.
The initial line lists were extracted from VALD. 
The total number of spectral lines, including lines
originating from the predicted levels, was more than 21.6 million for the
spectral range between 50 and 100\,000\,\AA. This line list was used for the
preselection procedure in the \llm\ code using the selection threshold
$\varepsilon=1$\%. Preselection allowed us to decrease the number of spectral lines
involved in the line blanketing calculation to about 300\,000--800\,000
depending on the model atmosphere parameters.

To calculate a Zeeman splitting of a spectral line one has to know the Land\'e
factors of the lower and upper atomic levels of the respective transition. The
VALD database does not provide Land\'{e} factors for all lines. We found that
roughly 4--10\% of the preselected spectral lines lack information about the Land\'e
factors. For the light elements, from He to Sc, the LS coupling approximation
is usually sufficient and was employed where necessary using the term designation
provided for each line in the VALD line lists. This approach 
reduced the number of lines with missing Land\'e factors to 1--4\%. For these
remaining lines without a proper term designation in the VALD lists
we assumed a classical Zeeman triplet splitting pattern with the
effective Land\'e factor $g_{\rm eff}=1.2$ (which is the average value for
the spectral lines with computed or experimental Land\'e factors). 
Further information on the
statistics of the considered spectral line lists is presented in
Table~\ref{number_lines}.

\subsection{Zeeman effect in the line opacity}

In the presence of a magnetic field an atomic level $k$ defined by the  quantum
numbers $J_k$, $L_k$, $S_k$ splits into the $2J_k+1$ states with the magnetic
quantum numbers $M_k=-J_k, \dots, +J_k$. The absolute value of the splitting is
defined by the field modulus $|\emph{\textbf{B}}|$ and by the Land\'e factor $g_k$,
which in the case of LS coupling can be calculated as
\begin{equation}
g_k=\frac{3}{2}+\frac{S_k(S_k+1)-L_k(L_k+1)}{2J_k(J_k+1)}.
\end{equation}
According to the selection rules the following transitions are allowed between
the splitted upper $u$ and lower $l$ levels
\begin{equation}
\Delta{M}=M_u-M_l=\left\{
\begin{array}{r}
+1\equiv b\\0\equiv p \\-1\equiv r\\
\end{array}.
\right.
\end{equation}
For a normal Zeeman triplet the subscript $p$ corresponds to the unshifted $\pi$
component, whereas $b$ and $r$ correspond to the blue- and red-shifted $\sigma$
components respectively. In the general case of an anomalous splitting pattern
the indices $p$, $b$, $r$ refer to the groups of the $\pi$ and $\sigma$
components.

The wavelength shift of the component $i_j$ ($j=p,b,r$,
$i_{p,b,r}=1, \dots, N_{p,b,r}$; $N_{p,b,r}$ is the number of components in
each group) relative to the laboratory line centre $\lambda_0$ is defined by
\begin{equation}
\Delta\lambda_{i_j}=\frac{e\lambda_0^2|\vec{B}|}{4\pi mc^2}\,(g_lM_l-g_uM_u)_{i_j}\,.
\label{lambda_shift}
\end{equation}
The relative strengths $S_{i_j}$ of the $\pi$ and $\sigma$ components 
are given by \citet{sobelman} and are listed in Table~\ref{splitting}.

\begin{table}
\caption{Relative strengths of the Zeeman $\pi$ and $\sigma$ components. Here 
$\Delta J=J_u-J_l$ and $M_i=-J_u,\dots,+J_u$. In the Zeeman regime considered in
our paper symmetric components have the same strength, 
e.g. $S_{i_r}=S_{i_b}$. \label{splitting}}
\begin{tabular}{cll}
\hline
\hline
$\Delta J$ & $S_{i_p}$ & $S_{i_{b,r}}$ \\
\hline
$\phantom{+}0$ & ${M_i^2}$ & $1/4 (J_u-M_i)(J_u+1+M_i)$ \\
          $+1$ & $J_u^2-M_i^2$ & $1/4 (J_u-M_i)(J_u-1-M_i)$ \\
          $-1$ & $(J_u+1)^2-M_i^2$ & $1/4 (J_u+1+M_i)(J_u+M_i+2)$ \\
\hline
\end{tabular}
\end{table}

In general, to calculate a stellar spectrum in the presence of a magnetic field one
has to consider the polarized radiative transfer equation for the  Stokes \emph{IQUV}
parameters. A solution of this problem requires special numerical techniques and is
very computationally expensive \citep[e.g.][]{piskunov02} and, hence, cannot be
easily included in the routine model atmosphere calculation of magnetic stars.
Nevertheless, the problem of the magnetic line blanketing can be simplified
considerably by assuming that the magnetic field vector is oriented perpendicular to
the line of sight, i.e. magnetic field is horizontal. In this case the effects
produced by the polarized radiative transfer on the Stokes $I$  are minimal and we can
use the transfer equation for the non-polarized radiation, treating individual Zeeman
components as independent lines. This approximation was used for the spectrum
synthesis analysis of individual lines by \citet{takeda} and \citet{nielsen}. The
method is fully justified for weak lines \citep{stenflo} and produces reasonable
results for moderate and strong spectral features, although neglecting the polarized
radiative transfer overestimates the magnetic intensification for the strongest
spectral lines. At the same time, given the complex and diverse angular dependence  of
the magnetic intensification of lines with different Zeeman patterns \citep{stift}, it
is difficult to estimate the bias introduced by restricting our modelling to a
horizontal field instead of  dealing with a full range of the surface field
orientations typical of  magnetic CP stars.

We modified the original line list by inserting additional spectral lines which 
correspond to individual Zeeman components of the anomalous splitting patterns. The
wavelength shifts were calculated using Eq.~(\ref{lambda_shift}) and the oscillator
strengths were established from the $gf$ values of the original lines using the
following formulas
\begin{equation}
\begin{array}{l}
(gf)_{i_p} = \displaystyle\frac{1}{2} S_{i_p}(gf), \vspace{0.1cm}\\
(gf)_{i_{b,r}} = \displaystyle\frac{1}{4} S_{i_{b,r}} (gf),
\end{array}
\end{equation}
where the sum of the relative strengths is 
normalized to unity for each group of the Zeeman components
\begin{equation}
\sum\limits_{i=1}^{N_p}S_{i_p}=\sum\limits_{i=1}^{N_b}S_{i_b}=\sum\limits_{i=1}^{N_r}S_{i_r}=1.
\end{equation}
The last column in Table~\ref{number_lines} shows the total number of 
the Zeeman components included in our line blanketing calculations. 

\begin{figure*}
\filps{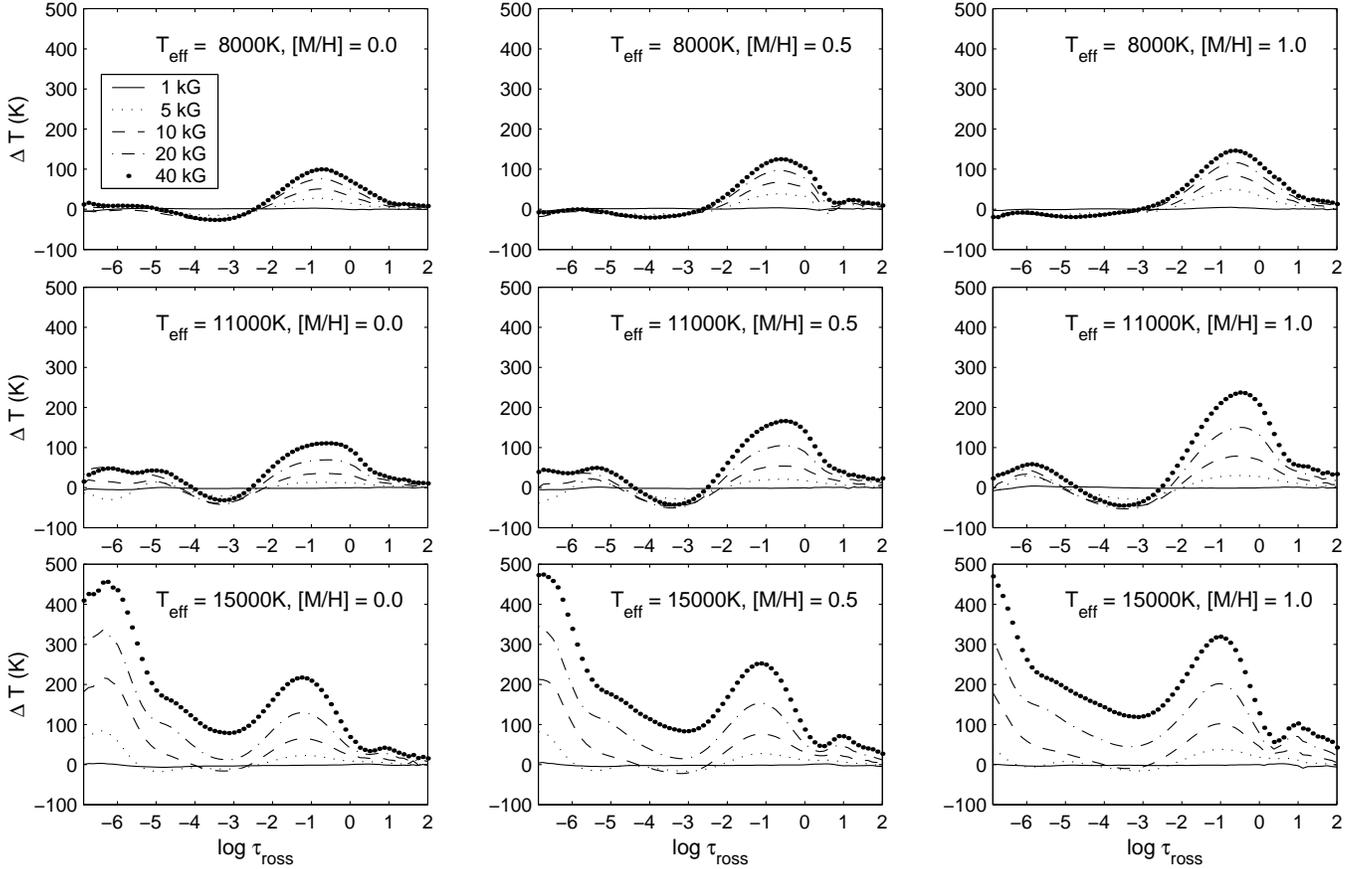}
\caption{Difference of temperature between magnetic and non-magnetic model 
atmospheres for the effective temperatures ${\tef=8000}$\,K, 
11\,000\,K, 15\,000\,K and metallicities ${[M/H]=0.0}$, +0.5, +1.0.}
\label{temperature}
\end{figure*}

\begin{figure*}
\filps{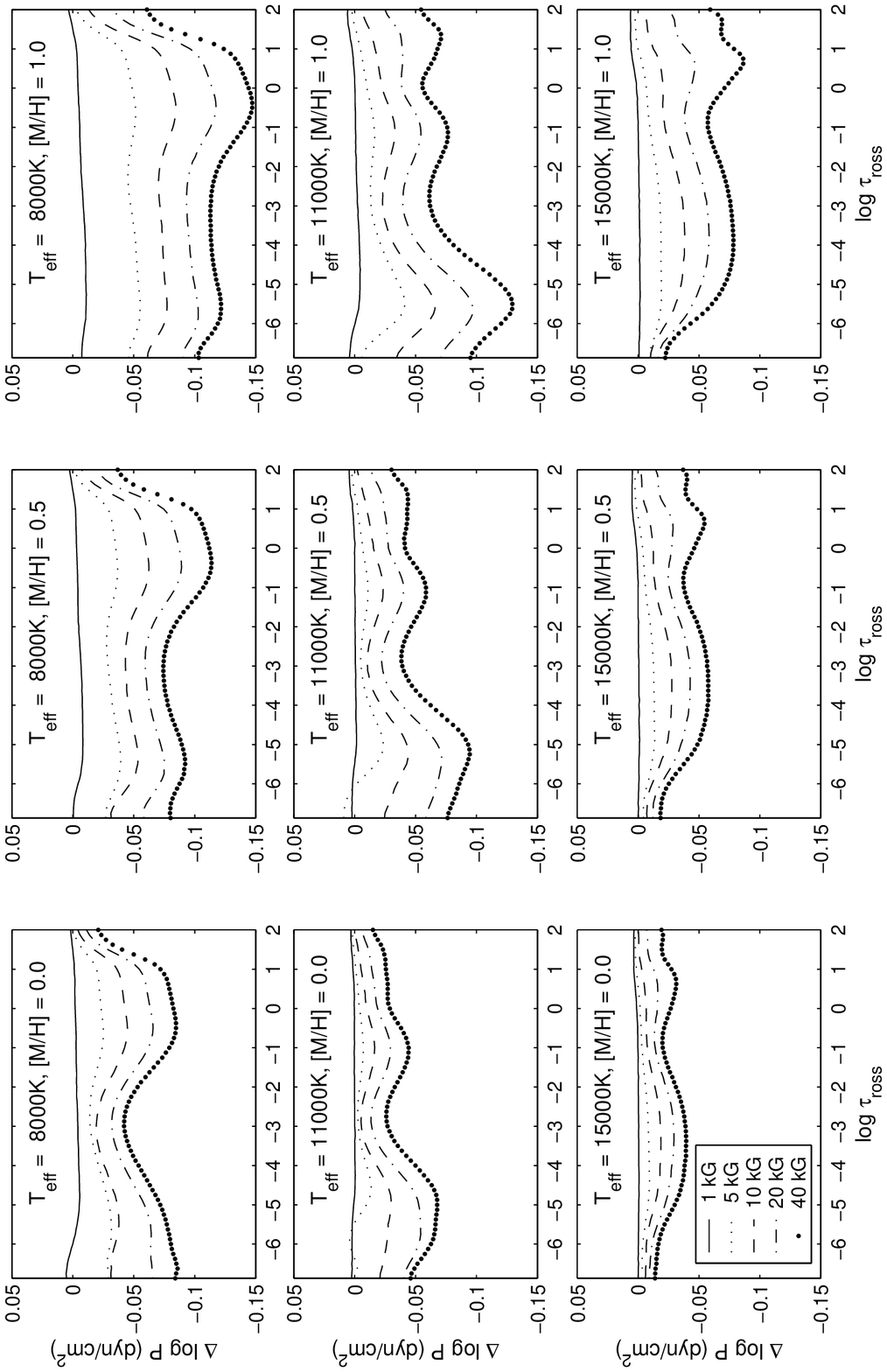}
\caption{Difference of pressure between magnetic and non-magnetic model 
atmospheres for the effective temperatures ${\tef=8000}$\,K, 
11\,000\,K, 15\,000\,K and metallicities ${[M/H]=0.0}$, +0.5, +1.0.}
\label{pressure}
\end{figure*}

\section{Numerical results}
\label{results}

Using the \llm\ code and treating the magnetic opacity as detailed above, we
have calculated a set of model atmospheres with the effective temperature
${\tef=8000}$\,K,  11\,000\,K, 15\,000\,K, surface gravity $\log g=4.0$,
metallicity $[M/H]=0.0$, +0.5, +1.0 and magnetic field 
strength 0, 1, 5, 10,
20 and 40\,kG. For a few models with $[M/H]=+1.0$ we also considered the enhanced
chromium abundance of $[Cr]=+2.0$ typical for some of the Cr-peculiar stars. 
The model atmosphere grid investigated in this paper covers substantial part of
the stellar parameter space occupied by the magnetic Ap and Bp stars.

The model atmosphere calculations were carried out on the Rosseland optical 
depth grid spanning from $+2$ to $-6.875$ in $\log\tauros$ and subdivided into
72 layers.  Convection was neglected, since it is generally believed that
strong global magnetic fields inhibit turbulent motions in the atmospheres of
peculiar stars. Due the assumed absence of convection and because of a direct
inclusion of the magnetic intensification in the modelling of spectral lines, we
adopted zero microturbulent velocity in all calculations with magnetic field.
A possibility to use pseudo-microturbulent velocity to mimic the effects of
magnetic field on the line opacity was investigated by computing several model
atmospheres with zero field and microturbulence in the range between 1 and 8
\kms.

Numerical tests demonstrated that a sufficient accuracy of the flux integration is
achieved by using the following wavelength ranges for the spectrum synthesis: from
500\,\AA\ to 50\,000\,\AA\ for ${\tef=8000}$\,K, from 500\,\AA\ to
30\,000\,\AA\ for ${\tef=11\,000}$\,K and from 100\,\AA\ to 30\,000\,\AA\ for
${\tef=15\,000}$\,K. For all models we used a wavelength step of 0.1\,\AA,
which resulted in the total number of frequency points in the range between
295\,000 and 495\,000.

At the first stage of creating models for each of the studied $\tef$, $[M/H]$
pairs we have used a standard \atlas\ \citep{kurucz13} model atmosphere for the
spectral line preselection and as an initial guess of the model structure. The
line selection threshold was set to 1\%. Experiments with the smaller selection
cutoffs in the range from $10^{-4}$\% to 0.5\% showed that, whereas including a
larger number of lines results in a considerable increase of computing time, it
leads to no more than 20--30\,K difference of the temperature in the surface
layers.

Two different initial approximations were investigated for calculation of 
magnetic model atmospheres. The first one uses a model atmosphere with zero
magnetic field, whereas in the second one a new model is calculated using a
converged model with a weaker field strength (i.e. the 0\,kG model is used as an
initial guess for the 1\,kG model, then the 5\,kG model is calculated starting from
the 1\,kG model, etc.). We found that the results of the two approaches are identical,
but the second one requires far less computing time due to a faster
convergence. Consequently, this iterative approach was used for the calculation of
the magnetic model atmosphere grid presented here.

In the following sections we study the influence of the magnetic line
blanketing on the model temperature and pressure structure and then investigate
the effects on the common photometric and spectroscopic observables: spectral
energy distribution, photometric colors in the Str\"omgren, Geneva and 
$\Delta a$ systems and profiles of the hydrogen Balmer lines.

\subsection{Model structure}

In Fig.~\ref{temperature} we show the difference between the temperature of
magnetic and non-magnetic models as a function of the optical depth. The magnetic line
blanketing typically increases temperature in the line forming region. This
distortion becomes more pronounced in a stronger field and higher metal
abundance. The most persistent effect of including magnetic field in the line
opacity calculation is the heating of atmosphere in the optical depth range
from $\log\tauros=-2$ to 1. This phenomenon appears due to the enhanced line
blanketing resulting from the additional absorption in Zeeman components. The
additional opacity leads to a redistribution of the absorbed energy back to the
lower atmospheric layers and, thus, to heating of the $\log\tauros \approx -2
\div 1$ region. This effect is present for all the models shown in
Fig.~\ref{temperature}.

In contrast, the behaviour of the upper layers depends strongly on the model parameters.
For coolest stars considered here (${\tef=8000}$\,K) we obtain a decrease of
temperature in the upper layers. For the ${\tef=11\,000}$\,K models this cooling is
limited to a small range of depths centred at $\log\tauros\approx-3.5$, whereas the
magnetic models computed for ${\tef=15\,000}$\,K display a significant increase of
temperature throughout the atmosphere. This behaviour is explained by the shift from
visual to UV of the wavelength interval which plays the most important role for the
atmospheric energy balance. For the hotter models the line density in UV becomes very
high due to Zeeman splitting and, therefore, even relatively high layers become
opaque and are able to absorb additional incoming radiation and contributes to blanketing
effect.

The pressure difference between magnetic and non-magnetic model atmospheres is
shown in Fig.~\ref{pressure}. The pressure stratification generally follows the
tendencies of the temperature distribution: deviations from the non-magnetic models
become larger with higher metal abundance and stronger magnetic field. The
pressure tends to decrease when magnetic effects on the line opacity are taken
into account. This is a consequence of a non-linear response of the model
structure to the modification of the line opacity and the respective temperature
increase due to the anomalous blanketing. 

\subsection{Energy distribution}

\begin{figure*}
\fifps{8.9cm}{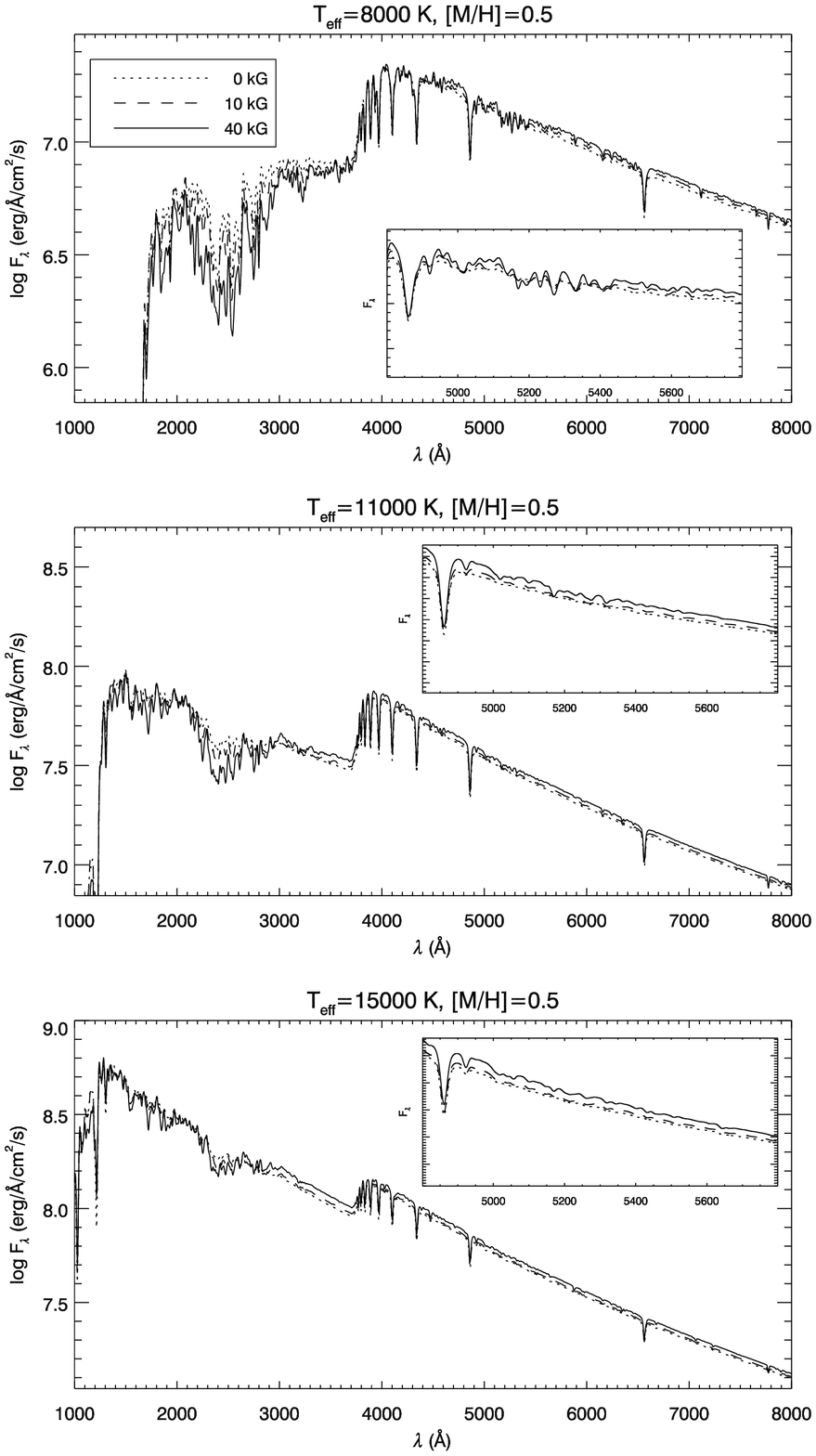}\nobreak
\fifps{8.9cm}{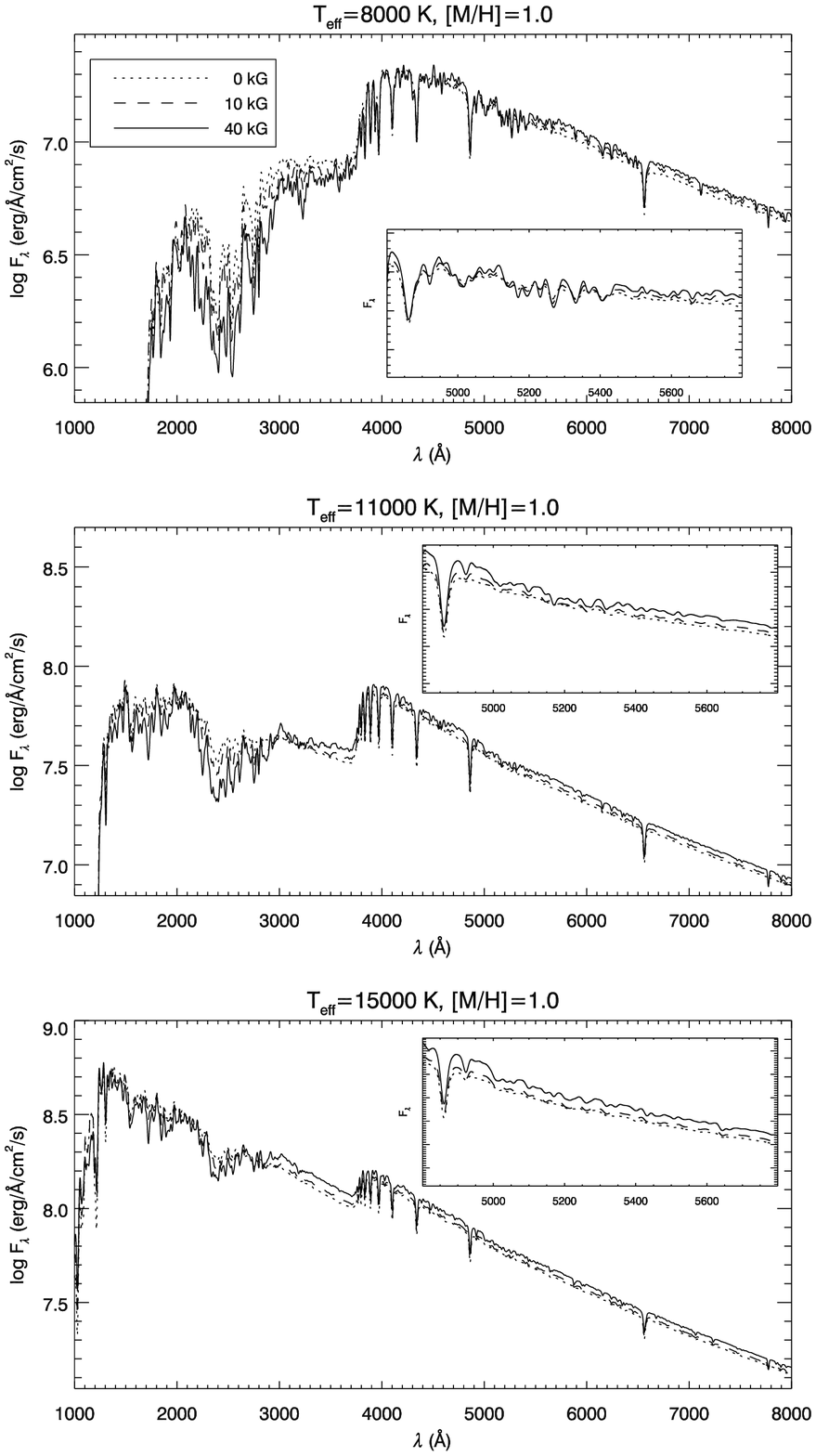}
\caption{Synthetic energy distributions from UV to near IR region
for effective temperatures ${\tef=8000}$\,K, 11\,000\,K, 15\,000\,K and 
metallicities $[M/H]=+0.5$, $[M/H]=+1.0$. To enable visual comparison of the 
different curves the original \llm\ stellar fluxes were convolved with
a Gaussian profile with $FWHM=15$~\AA. The insets show energy distributions
in the 5200\,\AA\ region.}
\label{energy}
\end{figure*}

The spectral energy distributions for the three values of effective temperature and
metal overabundance of $[M/H]=+0.5$ and $[M/H]=+1.0$ are presented in
Fig.~\ref{energy}. The theoretical energy distributions for magnetic stars are
compared with the calculation for the reference non-magnetic models. 

Three main anomalies are evident in the computed flux distribution of magnetic
stars. The most conspicuous signature of the magnetically modified line
blanketing is a flux deficiency in the ultraviolet spectral region and the
respective flux excess in the visual. The magnitude of the UV deficiency is a
moderate function of the effective temperature and clearly increases with the
magnetic field strength and metallicity.

The presence of a magnetic field changes the stellar flux distribution in opposite
direction in the visual and UV regions. At short wavelengths magnetic star
appears to be cooler in comparison with a non-magnetic object with the same
fundamental parameters, whereas in the visual magnetic star mimics a hotter
normal star.  The ``null wavelength'' where flux remains constant
progressively shifts to bluer wavelengths as the stellar effective temperature
increases.

Finally, we find that some of the theoretical flux distributions of magnetic
stars exhibit depression in the 5200\,\AA\ region. This spectral feature has a
well-known counterpart frequently observed in the spectra of peculiar stars
\citep{kupka1}. In
our theoretical calculations the 5200\,\AA\ depression is prominent at lower
$\tef$ but becomes rather small for hotter models.  The magnitude of the
depression increases with the magnetic field intensity and metal content of the
stellar atmosphere. 

\begin{figure*}
\filps{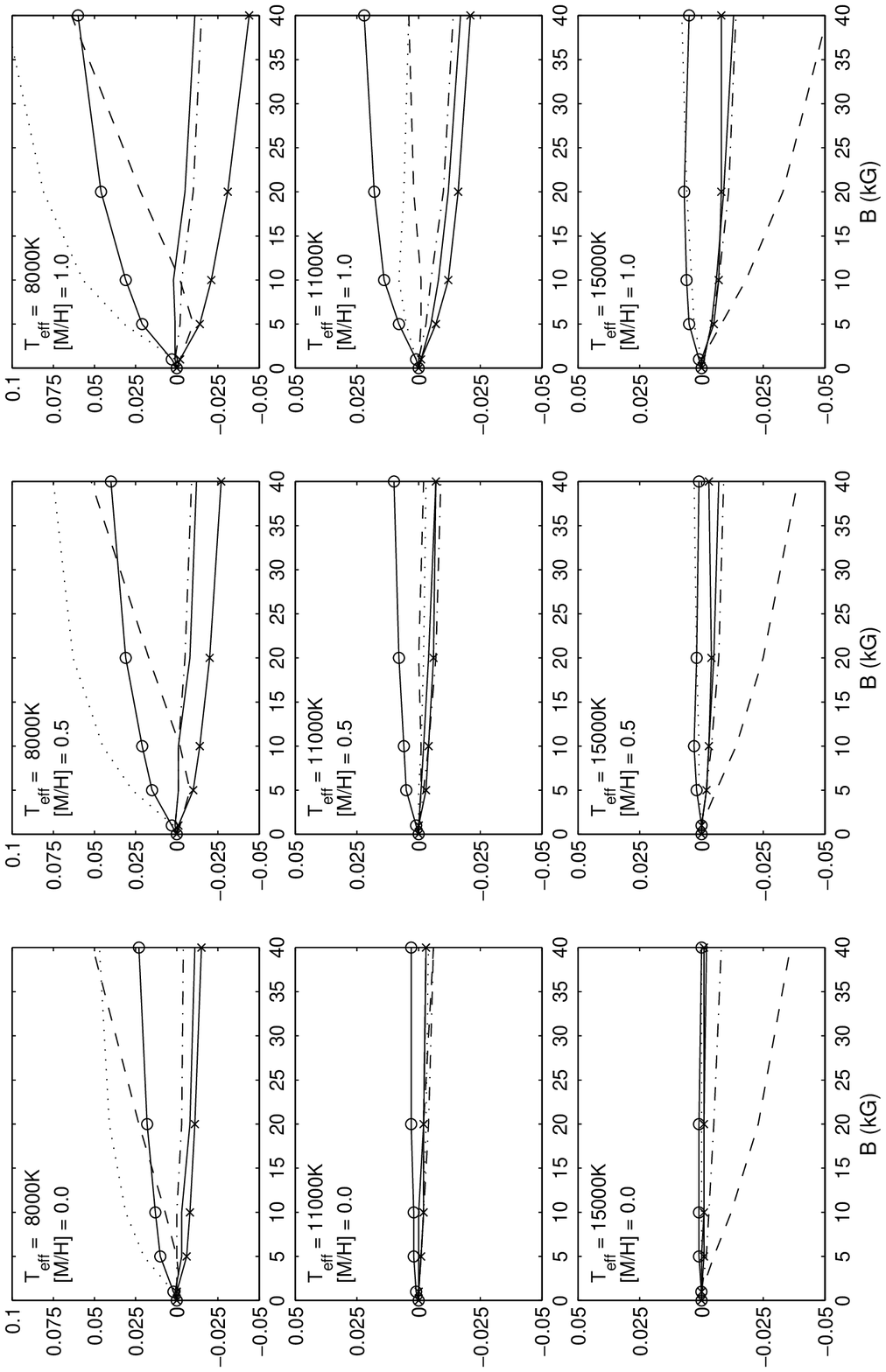}
\caption{Difference between photometric color indices of magnetic stars in
comparison with the predictions for the non-magnetic reference models. Effect of
magnetic field on the photometric parameters of the Str\"omgren system is shown
for $b-y$ (solid line), $m_1$ (dotted line), $c_1$ (dashed line) and $H\beta$
(dash-dotted line). The peculiarity indicators in the Geneva system, 
$\Delta(V_1-G)$ and $\Delta Z$, are shown
with open circles and crosses respectively.}
\label{colors}
\end{figure*}

\subsection{Colors}
\label{photom}

\begin{figure*}[t]
\figps{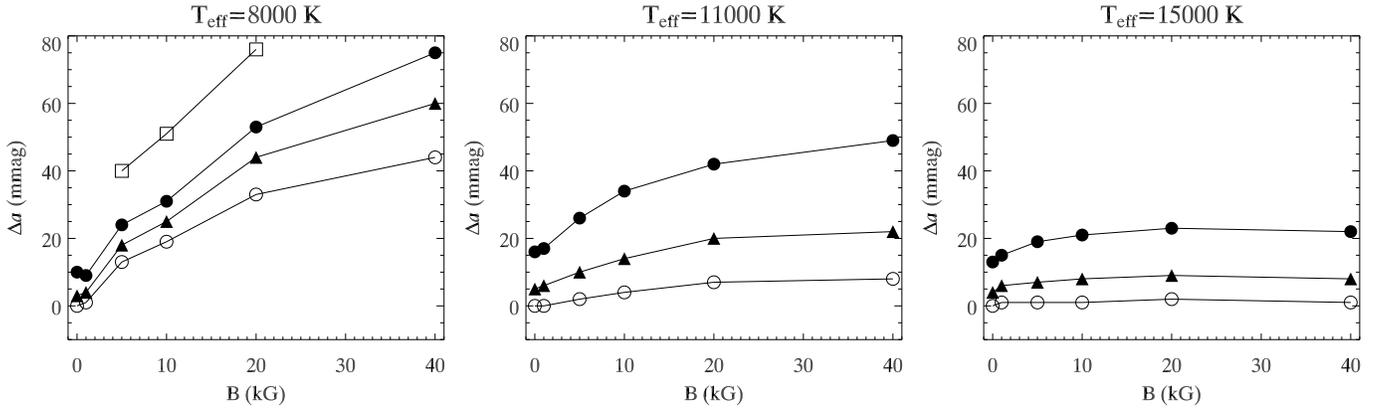}
\caption{The magnitude of the $\Delta a$ photometric index as a function of the
magnetic field strength for $\tef=8000$, 11\,000 and 15\,000~K. Different curves
show calculations with metal abundances $[M/H]=0.0$ (open circles),
$[M/H]=+0.5$ (triangles) and $[M/H]=+1.0$ (filled circles). Open squares
illustrate the effect of increasing Cr overabundance to $[Cr]=+2.0$ in the
$\tef=8000$, $[M/H]=+1.0$ models.}
\label{delta_a}
\end{figure*}

We studied the influence of the magnetic line blanketing on the  photometric
colors in the Str\"omgren $uvbyH\beta$, Geneva and  $\Delta a$ systems. The
dependence of various photometric parameters on magnetic field strength is
shown in Figs.~\ref{colors} and \ref{delta_a}.

All colors were calculated using modified computer codes by \citet{kurucz13},
which take into account transmission curves of individual photometric filters,
mirror reflectivity and a photomultiplier response function. In contrast to the
Kurucz's procedures \citep{relyea} which are based on the low-resolution
theoretical fluxes, our synthetic colors are computed from the energy
distributions sampled every 0.1\AA, so integration errors are expected to be
small.

The peculiar parameter $\Delta a$ was introduced by \citet{maitzen} to
measure the strength of the 5200\,\AA\ depression. Our calculations of the
synthetic $\Delta a$ photometry were carried out using the tabulated filter
transmission curves and the photomultiplier response functions as discussed in the
recent paper by \citet{kupka1}. In addition, the peculiarity parameters $Z$ and
$\Delta(V_1-G)$ of the Geneva photometric systems were computed according to
\citet{cramer1}.

As one expects, the behaviour of the photometric indices is closely related to the
flux redistribution between the visual and UV regions and the presence of several
flux depressions, such the one at 5200\,\AA. The shift of the ``null wavelength''
leads to alternation of the Balmer discontinuity amplitude as measured
by the $c_1$ index of the $uvbyH\beta$ system (see Fig.\,\ref{colors}). We
found that for all metallicities a relation between $c_1$ and the magnetic field
intensity is almost linear for ${\tef=8000}$\,K and ${\tef=15\,000}$\,K, with
positive and negative correlation respectively. On the other hand, for
${\tef=11\,000}$\,K the $c_1$ index does not show any trend with the magnetic field
strength because for this temperature the flux increases by a similar amount on 
both sides of the Balmer jump.

A relation between changes of all photometric indices and the intensity
of magnetic field depends strongly on the stellar effective temperature. For
low $\tef$ photometric changes are very pronounced, whereas for hotter magnetic
stars modification of the $uvbyH\beta$ photometric observables is fairly small
(except $c_1$). The Geneva photometric colors show a similar decrease of the
sensitivity to the magnetic effects with increasing $\tef$. This picture agrees
with the general temperature behaviour of the theoretical flux 
distributions evident in Fig.~\ref{energy}. 

The $\Delta a$ photometric system, characterizing the strength of the 5200~\AA\
depression, has long been considered as one of the most useful peculiarity
indicators for stars with non-solar chemical composition and strong magnetic
field (e.g., see Kupka et al. \citeyear{kupka1}, \citeyear{kupka2} and
references therein). Results of our calculations of the synthetic $\Delta a$
parameter are summarized in Fig.\,\ref{delta_a} and confirm its sensitivity to
the metal abundance and magnetic field strength. However, similar to the effects
observed for other medium and narrow-band photometric indicators, $\Delta a$ is
most influenced by magnetic field at lower $\tef$, whereas the trend of 
$\Delta a$ vs. the field strength quickly saturates for higher temperatures. 
For example, for $\tef=8000$\,K and $[M/H]=+1.0$ the $\Delta a$ 
index shows 0.043~mag difference between the 0\,kG and 20\,kG field strengths,
for $\tef=11\,000$\,K this difference decreases to 0.026~mag and it drops to
0.010~mag for the models computed with $\tef=15\,000$\,K.

The prominent temperature sensitivity of the behaviour of the 
synthetic $\Delta a$ indices arises 
because the absorption in the 5200~\AA\ region is dominated by the \ion{Fe}{i}
and low excitation \ion{Fe}{ii} lines. These features become weaker with increasing
$\tef$, hence the depression is stronger and more sensitive to the field strength
in cooler magnetic stars.

All photometric indices become more sensitive to the magnetic effects with
increasing metallicity. This is most clearly visible for the $\Delta a$ index
computed for $\tef=8000$ and 11\,000~K models. The respective panels in
Fig.~\ref{delta_a} demonstrate that, for instance, the difference  between
$\Delta a$ values for the models with ${\tef=8000}$\,K, ${[M/H]=0.0}$ and 
${[M/H]=+1.0}$ increases from the initial 0.010~mag for zero field to 0.031~mag
for $B=40$~kG. This effect is even more dramatic for ${\tef=11\,000}$\,K where
the difference $\Delta a_{[M/H]=+1.0} - \Delta a_{[M/H]=0.0}$ grows from 0.016
to 0.041~mag.

An anomalous concentration of individual species with abundant spectral lines may
have important implications for the $\Delta a$ value. For the models with
scaled solar abundance the 5200~\AA\ depression is produced mainly by the Fe lines.
Contribution of the lines of other elements becomes significant only when the
respective species are strongly overabundant. As illustrated in
Fig.~\ref{delta_a}, increasing the Cr overabundance by a factor of 100 relative
to the sun results in 0.016--0.023~mag growth of the $\Delta a$ index for 
${\tef=8000}$\,K, ${[M/H]=+1.0}$ model and the field strength 5--20~kG.

\citet{hauck} has demonstrated that the $\Delta(V_1-G)$ index of the Geneva
system could be used as a  peculiarity parameter for Ap stars. We found that
the behaviour of $\Delta(V_1-G)$ is comparable to that of the $\Delta a$ index.
The maximum change of the $\Delta(V_1-G)$ parameter relative to the non-magnetic
case is 0.060~mag for ${B=40}$\,kG, ${\tef=8000}$\,K and ${[M/H]=+1.0}$.

The $X$, $Y$, $Z$ parameters introduced by \citet{cramer2} are linear 
combinations of the Geneva photometric indices. The $Z$ value was suggested to
be an indicator of chemical peculiarity and the strength of the surface magnetic
field \citep{north}. We found that the maximum change of $Z$ relative to the
non-magnetic case equals to $-0.044$~mag for ${\tef=8000}$\,K, ${[M/H]=+1.0}$
and ${B=40}$\,kG. However, at higher temperatures $Z$ becomes less sensitive to
the magnetic field effects compared to $\Delta(V_1-G)$ and $\Delta a$.

Finally, we note that none of the photometric indicators of magnetic stars
proposed in the literature shows a linear trend over the whole range of the
considered magnetic field strength. Relations between changes of the photometric
indices and the magnetic field intensity show saturation effects for ${B\gtrsim
10}$\,kG. This saturation can be offset to stronger fields by increasing the
number of spectral lines affected by the Zeeman splitting and intensification,
which explains greater magnetic sensitivity of the synthetic photometry
computed for the models with lower $\tef$ and higher metal abundance.

\subsection{Hydrogen line profiles and metal lines}

\begin{figure*}[t*]
\figps{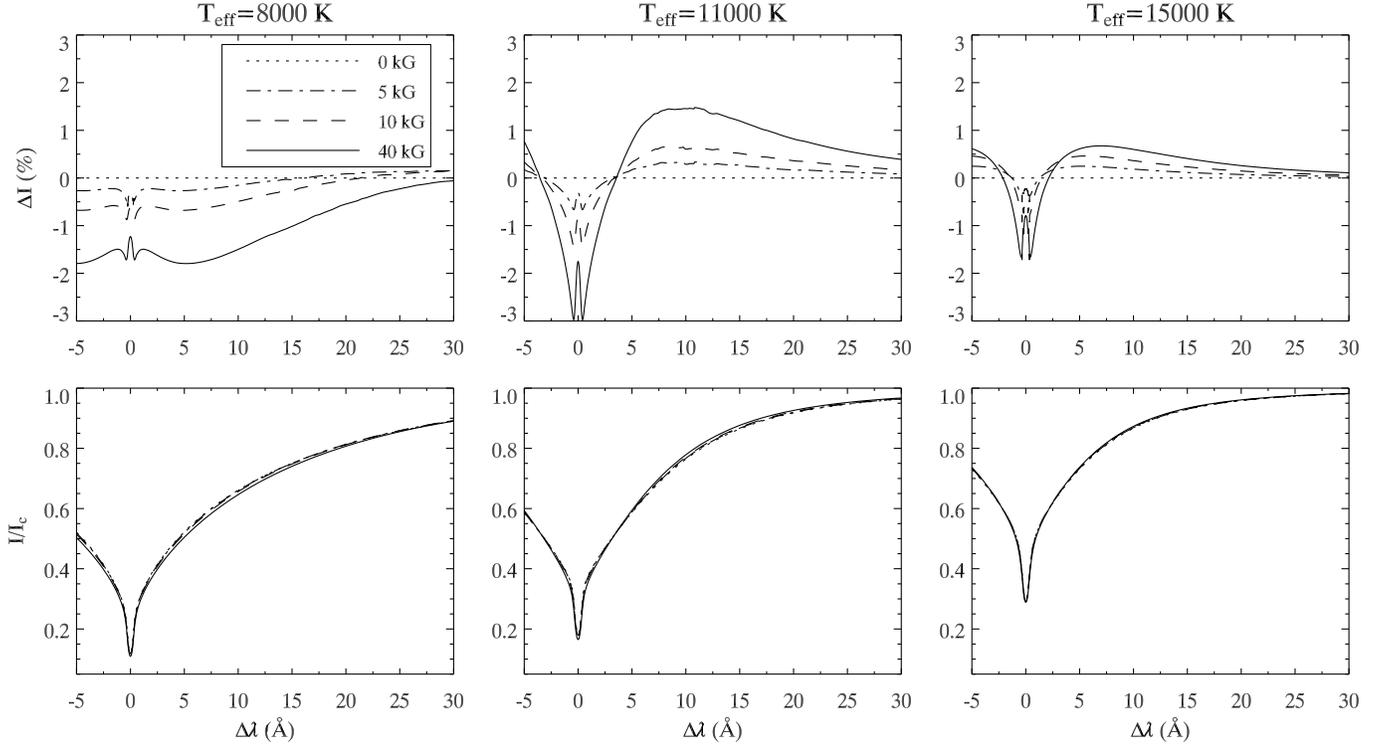}
\caption{Comparison between the synthetic H$\beta$ profiles computed for 
${[M/H]=+1.0}$ and different values of the magnetic field strength and $\tef$.
The lower panels show normalized profiles, whereas the upper plots illustrate
the difference between H$\beta$ calculated for the models with substantial 
magnetic line blanketing and the reference non-magnetic model.}
\label{hydrogen}
\end{figure*}

Profiles of the H$\alpha$, H$\beta$ and H$\gamma$ hydrogen Balmer lines were calculated
using the {\sc Synth} program \citep{piskunov}. This spectrum synthesis
code incorporates the new hydrogen line Stark broadening computations by
\citet{stehle}, as well as the improved treatment of the hydrogen line
self-broadening developed by \citet{barklem}. The comparison between the magnetic
and non-magnetic profiles of the H$\beta$ line calculated for models with 10
times the solar metal abundance is presented in Fig.~\ref{hydrogen}.

We found that the changes in the atmospheric structure due to the magnetic line
blanketing do not have a strong influence on the hydrogen line profiles. For
H$\beta$ the maximum change relative to the non-magnetic model amounts to about 3\%
of the  continuum level for ${B=40}$\,kG, but it does not exceed 1\% for
$B\le10$~kG, which is a more typical surface field strength for the majority
of magnetic CP stars. 

To estimate an effect of the anomalous atmospheric structure of the magnetic models
on the profiles of individual metal lines we computed synthetic spectra in the
4000--6000~\AA\ region for the zero field and the 10~kG field models and $[M/H]=+1.0$.
Since at this point we are interested to study only the effects of the model structure,
the spectrum synthesis calculations were carried out disregarding magnetic field. We
found that the depth of absorption lines is modified in the spectra computed with the
magnetic models. The strongest effect is found for $\tef=8000$\,K, where  in the presence
of the 10~kG field many medium strength and weak lines become sytematically shallower by
4--6\% of the continuum level relative to the calculation with the zero field model.
This discrepancy is reduced to less than 2--3\% for the $\tef=11\,000$ and 15\,000\,K
models. For these hotter atmospheres the difference between the line profiles computed
for the magnetic and non-magnetic models can be both negative and positive. For many
lines the main discrepancy is concentrated in the line wings, with the magnetic models
predicting slightly narrower line profiles. 

\subsection{Stellar atmospheric parameters and bolometric correction}

The overall modification of the energy distribution of magnetic stars as well as
the presence of localized absorption features may bias photometric  determination
of the stellar atmospheric parameters. As a part of the investigation in this paper, we
verified an influence of the magnetic line blanketing on the photometric determination
of $\tef$ and $\log g$. 

The {\sc TempLogg} code \citep{templogg} was applied to the
synthetic $uvbyH\beta$ colors calculated as described in Sect.~\ref{photom}. 
We found that for ${\tef=8000}$~K the maximum difference  between $\tef$ and $\log
g$ obtained for non-magnetic models and the corresponding parameters for the
${B=40}$\,kG models is $-190$\,K and $-0.27$~dex respectively. For the
${\tef=11\,000}$\,K models the temperature discrepancy is maximal ($+$86\,K) for
normal composition and decreases with increasing metal abundance, whereas the
difference of $\log g$ peaks at $-0.11$~dex. For the hottest models
($\tef=15\,000$\,K) the $\tef$ difference is $+541$~K, but $\log g$ is not
affected. It is clear that, even in the case of extreme magnetic field strengths,
the bias in determination of the fundamental parameters using the Str\"omgren photometry
does not result in a systematic deviation of $\tef$ and $\log g$ beyond the usual
error bars assigned to the photometrically determined stellar parameters.

Somewhat different results emerged from our analysis of the stellar parameters
determined using the Geneva photometric system. Effective temperature was derived
applying the code of \citet{kuenzli} to the synthetic Geneva colors and was found to  increase
systematically with the field strength for all three considered  $\tef$ values. The bias
introduced by magnetic field reaches 100--200~K for  ${\tef=8000}$--11\,000~K and 900~K
for ${\tef=15\,000}$~K. The surface gravity is modified by up to 0.2~dex. Thus, a
higher sensitivity to the metallic line absorption makes the standard Geneva
photometric calibration a less robust method for determination of the atmospheric
parameters of magnetic stars compared to the calibration in terms of the $uvbyH\beta$
colors.

An increase of the line blanketing due to magnetic effects and overabundance of metals
lead to a redistribution of the stellar flux from UV to visual. Consequently, magnetic CP
stars appear brighter in $V$ compared to normal stars with the same fundamental
parameters. Using the theoretical flux distributions calculated with the \llm\ code we
studied the effect of magnetic field and metallicity on the bolometric correction
(BC). For all stellar models this parameter decreases relative to the non-magnetic case by
up to 0.054--0.089~mag for $B=40$~kG and by no more than 0.025~mag for $B\le10$~kG.
The BC for the ${\tef=15\,000}$~K models is the most sensitive to magnetic field. At
the same time, even in the absence of the field, an increase of metal abundance leads
to a substantial modification of BC. It decreases by 0.079--0.115~mag when metal
abundance is changed from solar to $[M/H]=+1.0$.

\subsection{Enhanced microturbulence models}

Until the present investigation the most common approach to account for the
magnetic line blanketing in the model atmosphere analyses of magnetic CP stars was
to use an increased value of the microturbulent velocity \citep{muthsam,kupka2}.
For the magnetic field $B$ such a pseudomicroturbulent velocity is computed 
according to the following equation (e.g., Kupka et al. \citeyear{kupka0}):
\begin{equation}
\xi_{\rm mag} = 4.66\times10^{-13} c\,\lambda g_{\rm eff} B,
\label{vmacro}
\end{equation}
where $c$ is the speed of light in \kms, the field strength $B$ is measured in gauss, 
$\lambda$ is the wavelength in \AA\ and $g_{\rm eff}$ is the effective 
Land\'e factor. In previous model atmosphere studies $\lambda\approx5000$~\AA\ and
$g_{\rm eff}=1.0$--1.2 were typically adopted, which gives $\xi_{\rm
mag}\approx4$ and 8~\kms\ for the magnetic field strength of 5 and 10~kG respectively.

Having introduced a more realistic implementation of the magnetic intensification
in modelling of stellar atmospheres, we are in the position to verify the
extent to which models with the magnetic pseudomicroturbulence are able to match
the properties of our more sophisticated but, admittedly, also more
computationally expensive models. To study the performance of the enhanced
microturbulence models, the model structure, energy distribution and synthetic
$\Delta a$ indices computed for zero field and $\xi_{\rm mag} = 1$--8~\kms\ were
compared with the respective parameters for the $B=5$ and 10~kG models from our
main grid.

It appears that the enhanced microturbulence is indeed able to partially mimic the
behaviour of magnetic models, such as the energy redistribution from UV to visual
and the heating in certain atmospheric layers. However, an exact quantitative match
requires adopting a different $\xi_{\rm mag}$ parameter for each considered
quantity. For instance, to reproduce the temperature structure and energy
distribution of the $\tef=8000$~K models one has to use (depending on
metallicity)  $\xi_{\rm mag} = 2.1$--2.6~\kms\ and 3.7--4.1~\kms\ for $B=5$ and
10~kG respectively, which is roughly a factor of two smaller microturbulence
than predicted by Eq.~(\ref{vmacro}). At the same time, matching the anomalous
absorption in the 5200~\AA\ region requires a 2--3~\kms\ higher value of
$\xi_{\rm mag}$. Furthermore, models' sensitivity to the magnetic line
blanketing  decreases with $\tef$ faster than the influence of microturbulence.
Consequently, for the $\tef=15\,000$~K models one has to adopt nearly twice as
weaker $\xi_{\rm mag}$ to simulate effect of the same field as in 
cooler models.

Results of the calculations in this section demonstrate that a properly chosen
value of $\xi_{\rm mag}$ can only be used as a very rough guess of the effects
due to magnetic field. However, it is evident that the models with enhanced 
microturbulence are not suitable for detailed analysis of magnetic stars and
cannot be used for modelling individual features of the stellar
flux distributions.

\section{Conclusions and discussion}
\label{discussion}

Construction of the model atmospheres of CP stars is considerably more complicated
in comparison to that of normal stars because such modelling has to take into account and
explain a range of phenomena not present in the atmospheres of normal stars. The
main and most important distinction of magnetic CP stars from the other Main
Sequence objects is the presence of a strong global magnetic field. It plays
a central role in producing chemical abundance anomalies and inhomogeneities 
and leads to a peculiar atmospheric structure, unusual line profiles
and an anomalous flux distribution.

The Zeeman splitting of spectral lines is one of the principal effects taking place
in the magnetic stellar atmosphere. In this study we have computed a grid of model
atmospheres of peculiar A and B stars accounting for the magnetic line blanketing.
Our model grid has covered the range of possible atmospheric parameters of CP
stars, which allowed us to analyze the behaviour of the model structure and  observed
characteristics for different magnetic field, abundances and effective
temperature. We have used the new model atmosphere code \llm\ and up-to-date
compilations of the atomic line data from the VALD database to insure an accurate
modelling.

A somewhat simplified ``horizontal" model of magnetic field  allowed us to
reduce computational costs by solving the radiative transfer  equation for the
non-polarized radiation. This approach does not fully reflect the influence of
magnetic field on the formation of spectral lines but, nevertheless, it represents
the most accurate treatment of the Zeeman effect in model atmosphere
calculation used so far.

Our models show that the enhanced line blanketing due to the magnetic intensification 
of spectral lines produces heating of the certain atmospheric layers and this
effect increases with effective temperature, magnetic field strength and metal
abundance.

We found that the model atmospheres with the magnetic line blanketing produce fluxes 
that are deficient in the UV region by ${0.2-0.3}$\,mag for ${B=10}$\,kG and
about ${0.3-0.8}$\,mag for ${B=40}$\,kG. Moreover, the presence of a magnetic
field leads to the flux redistribution from the UV to visual region due to backwarming.
This property of the theoretical models is in agreement with those observations of CP
stars \citep{leckrone3,molnar,jamar} which have revealed the presence of the flux
redistribution and demonstrated that the flux changes in the visible and
ultraviolet regions are inversely correlated.

The energy distributions of the calculated model atmospheres show that an increase
of  effective temperature shifts the ``null wavelength'' where flux  remains
unchanged to shorter wavelengths. The results of our modelling suggest that the
``null wavelength'' approximately falls at 3800~\AA\ for ${\tef=8000}$\,K, at
2900~\AA\ for ${\tef=11\,000}$\,K and at 2600~\AA\ for ${\tef=15\,000}$\,K and
slightly depends on metallicity and field strength. This fact is in 
concordance with the observational data and numerical experiments by
\citet{leckrone2} who examined enhanced opacity models and found that the
``null wavelength'' shifts toward UV when effective temperature is increased.

Magnetic CP stars demonstrate several flux depressions in the visual and 
ultraviolet region unlike other stars. The most prominent feature in the visual is
the flux depression centred at 5200\,\AA. Our numerical results show that the strong
feature near 5200\,\AA\ appears for lower effective temperature and  vanishes for
higher effective temperature and its amplitude increases with the magnetic field
strength and metallicity. Comparison of the synthetic $\Delta a$ indices with typical
values of this parameter observed for cool magnetic stars demonstrate that our
calculation are able to reproduce the whole range of the 5200~\AA\ depression
strength. Furthermore, our results argue that for modelling the 5200~\AA\ feature
in cool CP stars an accurate treatment of magnetic field is of no less importance
compared to using individual abundances \citep{kupka2}. A 10~kG magnetic field
changes the value of the $\Delta a$ parameter by as much as 20--30~mmag, which is
substantial even compared to the largest observed $\Delta a$ values.

Since, historically, photometric data are widely used for the stellar classification and
atmospheric parameter determination, we have calculated indices in the $uvbyH\beta$ and
Geneva photometric systems as well as the peculiar indices $\Delta a$, $Z$ and
$\Delta(V_1-G)$. We found that changes of most  photometric parameters due to the
influence of magnetic field are  noticeable for low effective temperatures, whereas
for hotter stars sensitivity to magnetic field is reduced considerably. This
behaviour of the theoretical models agrees with the flux distribution  properties and
with the study by \citet{cramer2}, who noted that the effects of magnetic  field on the
energy distribution become less important with an increase of temperature.

The peculiar indices $\Delta a$, $Z$ and $\Delta(V_1-G)$ are frequently used for 
identification and assessment of the properties (in particular the surface field
strength) of CP stars. We examined the influence of magnetic field on these indices. 
We found that the magnetic modification of the peculiar photometric parameters
is relatively small (maximum change of $\Delta a$ is 0.065~mag for 40\,kG field) and
the relation between their values and the magnetic field strength is not linear due to
the saturation effects for stronger field (${B\gtrsim 10}$\,kG). In any case, the
$\Delta a$ parameter appears to be the most sensitive to magnetic field, whereas
$Z$ and $\Delta(V_1-G)$ are less sensitive. \citet{cramer2} pointed
that the saturation of the photometric effects for large fields is an
observational fact and has to be explained by a model atmosphere analysis. Our
numerical results reproduce and provide a clear explanation of the behaviour of
the photometric indices.

Finally, we investigated the question to what extent the anomalous magnetic opacity
affects the photometric determination of the atmosphere parameters of CP stars.  We found
that, although an enhanced metal abundance and magnetically modified line opacity
produce some changes of the model atmosphere structure and the flux distribution,
these changes are not very large and do not result in a strong modification of the
optical photometric colors. Consequently, the model atmosphere parameters derived using
the photometric calibrations for normal stars are not far from their true values. Thus,
despite the presence of several local anomalies of the stellar energy distribution
and the overall flux redistribution from the UV to visual region, peculiar stellar
models do not appear to mimic standard models with a systematically different
$\tef$. Standard calibrations of the Str\"omgren photometry appear to be
especially robust for the determination of the parameters of magnetic stars.
Thus, we conclude that by itself the magnetic opacity does not introduce significant
errors in the photometric estimates of CP-star parameters.

 We also showed that the hydrogen Balmer line and absorption features of metals are
not very sensitive to the changes induced by the magnetic line blanketing on the model
atmosphere structure.  Although the effect on the overall flux distribution may be
noticeable, the changes in the depths of metal lines do not exceed a few \% of the
continuum level for the 10~kG field, which justifies neglecting magnetic line blanketing in
the routine abundance and spectrum synthesis studies of the majority of magnetic CP stars.


\begin{acknowledgements}
We are grateful to Prof. V.~Tsymbal for enlightening discussions.

This work was supported by the Lise Meitner fellowship to OK (FWF
project M757-N02), INTAS grant 03-55-652 to DS and by the Austrian {\em
Fonds zur F\"orderung der wissenschaftlichen Forschung} (P--14984), the
{\em BM:BWK} (project {\sc corot}). Financial support came from the {\em
RFBR} (grant 03-02-16342) and from a {\em Leading Scientific School},
grant 162.2003.02.
\end{acknowledgements}

\end{document}